\documentclass[11pt,notref,notcite]{article}
\usepackage{amsmath,amsthm}
\usepackage{amssymb}
\usepackage{cite}
\usepackage{indentfirst}
\newtheorem{theorem}{Theorem}[section]

\newcommand{\be}{\begin{equation}}
\newcommand{\ee}{\end{equation}}
\newtheorem{lemma}{Lemma}[section]
\numberwithin{equation}{section}
\def \K{[\![}
\def \J{]\!]}
\def \k{[\![}
\def \j{]\!]}

\title{Symmetries for the Ablowitz-Ladik hierarchy: II. Integrable discrete nonlinear Schr\"odinger equation
and discrete AKNS hierarchy
}
\author {Da-jun Zhang\footnote{Corresponding author. E-mail: djzhang@staff.shu.edu.cn},~~Shou-ting Chen
\\
{\small\it Department of Mathematics, Shanghai University, Shanghai
200444, P.R. China}}
\begin{document}
\maketitle

\begin{abstract}

In the paper we continue to consider symmetries related to the Ablowitz-Ladik hierarchy.
We derive symmetries for the integrable discrete nonlinear Schr\"odinger hierarchy and discrete AKNS hierarchy.
The integrable discrete nonlinear Schr\"odinger hierarchy are in scalar form
and its two sets of symmetries are shown to form a Lie algebra.
We also present discrete AKNS isospectral flows, non-isospectral flows and their recursion operator.
In continuous limit these flows go to the continuous AKNS flows
and the recursion operator goes to the square of the AKNS recursion operartor.
These discrete AKNS flows form a Lie algebra which plays a key role in constructing symmetries and their
algebraic structures for both the integrable discrete nonlinear Schr\"odinger hierarchy and discrete AKNS hierarchy.
Structures of the obtained algebras  are different structures from those in continuous cases
which usually are centerless Kac-Moody-Virasoro type.
These algebra deformations are explained through continuous limit and \textit{degree}
in terms of lattice spacing parameter $h$.

\vskip 5pt

\\
%\noindent{\bf MSC:}\quad 37K10
\end{abstract}

\section{Introduction}

In the previous paper \cite{ZDJ-AL-2010}, to which we refer as Part
I, we have derived  symmetries and their algebras for the
isospectral and non-isospectral four-potential Ablowitz-Ladik (AL)
hierarchies, and these results including symmetries of hiearachies
were shown to be reduced to the two-potential case.

Among the equations related to the two-potential AL spectral
problem, one of the most physically meaningful systems is the
integrable discrete nonlinear Schr\"{o}dinger(IDNLS)
equation\cite{AL-75-JMP,AL-76-JMP,Ablowitz-04-book}:
\begin{equation}
iQ_{n,t}=Q_{n+1}+Q_{n-1}-2Q_{n}+\varepsilon
|Q_{n}|^2(Q_{n+1}+Q_{n-1}), \label{idnlse}
\end{equation}
where $\varepsilon=\pm 1$ and $i$ is the imaginary unit. This
equation differs from the discrete NLS equation
\begin{equation}
iQ_{n,t}=Q_{n+1}+Q_{n-1}-2Q_{n}+2\varepsilon Q_{n}|Q_{n}|^2
\label{dnlse}
\end{equation}
by the nonlinear term, which is not integrable but arose in many
important physical contexts (cf. the introduction chapter of
\cite{Ablowitz-04-book} and the references therein).

With regard to symmetries, Refs.\cite{Levi-TMP-01,Levi-JNMP-03}
derived point and generalized symmetries for the IDNLS equation
\eqref{idnlse}. Two subalgebras were obtained\cite{Levi-TMP-01} and then a general
structure for the whole symmetry algebra was described\cite{Levi-JNMP-03}. In the present
paper we hope to go further than Refs. \cite{Levi-TMP-01,Levi-JNMP-03}. In fact,
Eq.\eqref{idnlse} consists of positive as well as negative order AL
isospectral flows, which corresponds to a central-difference
discretization for the second order derivative in the continuous NLS
equation. Since in Part I and also in \cite{ZDJ-06-PLA} we have derived
algebraic relations for arbitrary two flows among the positive and
negative-order, isospectral and non-isospectral, hierarchies related
to the two-potential AL spectral problem, it is possible to get
infinitely many symmetries with a clear algebraic structure for the
IDNLS equation \eqref{idnlse}.

Our plan is the following. We will start from the isospectral and
non-isospectral flows of the two-potential AL system, which serve as
basic flows. By suitable linear combinations we can get isospectral
and non-isospectral IDNLS flows separately (cf.
\cite{Ladik-77-JMP,Levi-TMP-01} where a uniformed hierarchy was
given). Since the IDNLS equation \eqref{idnlse} is in scalar form,
we then try to derive algebraic structure of the scalar IDNLS flows.
By the obtained structure we can present infinitely many symmetries
which form a Lie algebra with clear structure. Going further we can
get similar results for the IDNLS hierarchy. As in the continuous
AKNS system, the IDNLS flows can be a reduction of some flows which
we call the discrete AKNS (DAKNS) flows in the paper. We find these flows
and their recursion operator correspond to the continuous AKNS flows
and the square of the AKNS recursion operator. In our paper the algebraic
structure of the DAKNS flows will play a key role in deriving
symmetries and their algebras for the IDNLS hierarchy and the DAKNS
hierarchy. Finally, by comparison one can see the difference between
the obtained algebras and their continuous counterparts.
These algebra deformations will also be explained
by considering continuous limit and introducing \textit{degree} of discrete elements.

The paper is organized as follows. Sec.2 contains basic results of the two-potential AL system,
 including flows, recursion operator and algebraic structure.
In Sec.3 we present isospectral and non-isospectral IDNLS flows and
their algebraic structures. Sec.4 derives  symmetries and Lie
algebras for the IDNLS equation and hierarchy. Sec.5 discusses the
DAKNS flows and continuous limits. In Sec.6  algebra deformations
from discrete case to continuous case are listed out and explanation
follows. Finally, Sec.7 gives conclusions.

\section{Flows related to the AL spectral problem}

Let us list out the main results of Ref.\cite{ZDJ-06-PLA} as the
starting point. We will also follow the notations used in
\cite{ZDJ-06-PLA} without any confusion with Part I. The
two-potential AL spectral problem (also called discrete
Zakharov-Shabat spectral problem) is
\begin{equation}
\Phi_{n+1}=M\Phi_n, ~~M=\Biggl (
                  \begin{array}{cc}
                   \lambda    &  Q_n   \\
                  R_n  & \frac{1}{\lambda}
                 \end{array}
             \Biggr ),~~
     u_n =\Biggl(
           \begin{array}{c}
               Q_n\\
               R_n
          \end{array}
        \Biggr ), ~~
    \Phi_n= \Biggl (
           \begin{array}{c}
            \phi_{1,n}\\
            \phi_{2,n}
          \end{array}
         \Biggr ).
\label{AL-n}
\end{equation}
The isospectral AL hierarchy  and non-isospectral AL hierarchy are
respectively
\begin{subequations}
\label{AL-hie}
\begin{align}
&u_{n, t}=K^{(l)}=L^{l}K^{(0)}, \quad
K^{(0)}=(Q_{n}, -R_{n})^{T},\quad l\in \mathbb{Z},\label{AL-hie-i}\\
&u_{n, t}=\sigma^{(l)}=L^{l}\sigma^{(0)}, \quad
\sigma^{(0)}=(2n+1)(Q_{n}, -R_{n})^{T},\quad l\in
\mathbb{Z},\label{AL-hie-n}
\end{align}
\end{subequations}
where $L$ is the recursion operator defined as
\begin{equation}
L\!=\!\!\left(\!
\begin{array}{cc} E& 0\\0 & E^{-1}
\end{array}\!\!\right)
\!+\!\left(\!\!\!
\begin{array}{c} -Q_{n}E\\R_{n}
\end{array}\!\right)\!(E-1)^{-1}(R_{n}E, Q_{n}E^{-1})+\gamma_{n}^{2}
\left(\!\!\!
\begin{array}{c} -EQ_{n} \\R_{n-1}
\end{array}\!\!\right)\!(E-1)^{-1}(R_{n},
Q_{n})\frac{1}{\gamma_{n}^{2}}, \label{recur-op}
\end{equation}
$\gamma_{n}=\sqrt{1-Q_{n}R_{n}}$ and $E$ is a shift operator defined as $E^j f(n)=f(n+j),~\forall j\in
\mathbb{Z}$. $L$ is invertible
with
\begin{equation*}
L^{-1}\!=\!\!\left(\!
\begin{array}{cc} E^{-1}& 0\\0 & E
\end{array}\!\!\right)
\!+\!\left(\!\!\!
\begin{array}{c} Q_{n}\\-R_{n}E
\end{array}\!\right)\!(E-1)^{-1}(R_{n}E^{-1}, Q_{n}E)+\gamma_{n}^{2}
\left(\!\!\!
\begin{array}{c} Q_{n-1} \\-ER_{n}
\end{array}\!\!\right)\!(E-1)^{-1}(R_{n},
Q_{n})\frac{1}{\gamma_{n}^{2}}.
\end{equation*}

The isospectral flows $\{K^{(l)}\}$ and non-isospectral flows
$\{\sigma^{(l)}\}$ form a centreless Kac-Moody-Virasoro (KMV) algebra, $\forall l,s \in \mathbb{Z}$,
\begin{subequations}
\label{alg-K-sigma}
\begin{align}
\k K^{(l)}, K^{(s)} \j &=0,\label{alg-K-sigma-a}
\\
\k K^{(l)}, \sigma^{(s)} \j & =2l K^{(l+s)}, \label{alg-K-sigma-b} \\
\k \sigma^{(l)}, \sigma^{(s)} \j &=2(l-s)\sigma^{(l+s)},
\label{alg-K-sigma-c}
\end{align}
\end{subequations}
where the Lie product $\k \cdot, \cdot \j$ is defined as in Part I,
i.e.,
\begin{equation}
\K    f, g\J   =f^{\prime}[g]-g^{\prime}[f],
\end{equation}
in which $f=(f_1(u_n),f_2(u_n))^T, g=(g_1(u_n),g_2(u_n))^T$,
$f^{\prime}[g]$ is the Gateaux derivative of $f$ w.r.t $u_n$ in
direction $g$, i.e.,
\begin{equation}
f^{\prime}[g]=f(u_n)'[g]=\frac{d}{d
\epsilon}\Bigr|_{\epsilon=0}f(u_n+\epsilon g),  ~~\epsilon\in
\mathbb{R}, \label{def-gat}
\end{equation}
and vice verse for $g'[f]$. \eqref{alg-K-sigma} is a generalization of the results in \cite{Ma-Tamizhimani-JPSJ-1999}.

\section{Isospectral and non-isospectral IDNLS hierarchies}

\subsection{Hierarchies}

To derive IDNLS hierarchies, let us introduce auxiliary flows
\begin{subequations}\label{K-j-0}
\begin{align}
K^{(0)}_{[0]}&=K^{(0)},\\
K^{(0)}_{[1]}&=\frac{1}{2}(L-L^{-1})K^{(0)}=
\frac{1}{2}\left(\begin{array}{c}
       (1-Q_{n}R_n)(Q_{n+1}-Q_{n-1})\\
       (1-Q_{n}R_n)(R_{n+1}-R_{n-1})
\end{array}\right),
\end{align}
\end{subequations}
\begin{subequations}\label{sigma-j-0}
\begin{align}
\sigma^{(0)}_{[0]}=&\frac{1}{2}\sigma^{(0)},\\
\sigma^{(0)}_{[1]}=&\frac{1}{4}(L-L^{-1})\sigma^{(0)}=\frac{1}{4}
\left(\begin{array}{c}
       (1-Q_{n}R_n)[(2n+3)Q_{n+1}-(2n-1)Q_{n-1})]\\
       (1-Q_{n}R_n)[(2n+3)R_{n+1}-(2n-1)R_{n-1})]
\end{array}\right)\notag\\
&~~~~~~~~~~~~~~~~~~~~~~~-\frac{1}{2}\left(\begin{array}{c}
       Q_{n}\\
       -R_n
\end{array}\right)(E-1)^{-1}(Q_{n+1}R_{n}-Q_{n}R_{n+1}),
\end{align}
\end{subequations}
and  operator
\begin{align}
\label{L-AKNS} \mathcal{L}=&L-2I+L^{-1} \notag\\=&\left(
\begin{array}{cc} E+E^{-1}-2& 0\\0 & E+E^{-1}-2
\end{array}\right)+\gamma_{n}^{2}\left(
\begin{array}{cc} Q_{n-1}-Q_{n+1}E\\R_{n-1}-R_{n+1}E
\end{array}\right)(E-1)^{-1}(R_{n}, Q_{n})\frac{1}{\gamma_{n}^{2}}\notag\\
&+\left(
\begin{array}{cc} Q_{n}\\-R_{n}E
\end{array}\right)(E-1)^{-1}(R_{n}E^{-1},
Q_{n}E)-\left(
\begin{array}{cc} Q_{n}E\\-R_{n}
\end{array}\right)(E-1)^{-1}(R_{n}E,
Q_{n}E^{-1}),
\end{align}
where $I$ is the $2\times 2$ unit matrix. By $*$ we denote complex conjugate.
Then we define
\begin{subequations}
\begin{align}
\widetilde{K}^{(0)}_{[0]}&=K^{(0)}_{[0]}\Bigr|_{R_n=-\varepsilon Q_n^*}=(Q_n,\varepsilon Q_n^*)^T,\\
\widetilde{K}^{(0)}_{[1]}&=K^{(0)}_{[1]}\Bigr|_{R_n=-\epsilon
Q_n^*}=\frac{1}{2} \left(\begin{array}{r}
       (1+\varepsilon|Q_{n}|^2)(Q_{n+1}-Q_{n-1})\\
       -\varepsilon(1+\varepsilon|Q_{n}|^2)(Q^*_{n+1}-Q^*_{n-1})
\end{array}\right),
\end{align}
\label{K0-j}
\end{subequations}
\begin{subequations}
\begin{align}
\widetilde{\sigma}^{(0)}_{[0]}=\sigma^{(0)}_{[0]}\Bigr|_{R_n=-\varepsilon
Q_n^*}=
&(n+\frac{1}{2})(Q_n,\varepsilon Q_n^*)^T,\\
\widetilde{\sigma}^{(0)}_{[1]}=\sigma^{(0)}_{[1]}\Bigr|_{R_n=-\varepsilon
Q_n^*}=&\frac{1}{4}\left(\begin{array}{c}
       (1+\varepsilon|Q_{n}|^2)[(2n+3)Q_{n+1}-(2n-1)Q_{n-1})]\\
       -\varepsilon(1+\varepsilon|Q_{n}|^2)[(2n+3)Q^{*}_{n+1}-(2n-1)Q^{*}_{n-1})]
\end{array}\right)\notag\\
&  +\frac{1}{2}\varepsilon\left(\begin{array}{c}
       Q_{n}\\
       \varepsilon Q_{n}^{*}
\end{array}\right)(E-1)^{-1}(Q_{n+1}Q^{*}_{n}-Q_{n}Q^{*}_{n+1}),
\end{align}
\label{sigma0-j}
\end{subequations}
the  operator
\begin{equation}
\widetilde{\mathcal{L}}=\mathcal{L}|_{R_n=-\varepsilon Q_n^*},
\label{L-idnls}
\end{equation}
and further define the flows
\begin{subequations}
\begin{align}
\widetilde{K}^{(l)}_{[j]}&=\widetilde{\mathcal{L}}^l\widetilde{K}^{(0)}_{[j]},~~j\in\{0,1\},\\
\widetilde{\sigma}^{(s)}_{[j]}&=\widetilde{\mathcal{L}}^s
\widetilde{\sigma}^{(0)}_{[j]},~~ j\in\{0,1\}.
\end{align}
\label{idnlse-flows}
\end{subequations}
To make clear the relationship of two components in
$\widetilde{K}^{(l)}_{[j]}$ and $\widetilde{\sigma}^{(s)}_{[j]}$, we
introduce function sets
\begin{equation*}
\mathcal{A}_{[j]}(\varepsilon)=\{(f_1,(-1)^j\varepsilon
f_1^*)^T\},~~ j\in\{0,1\},
\end{equation*}
where $f_1$ is an arbitrary scalar function. Then one can find that
\begin{equation}
\widetilde{K}^{(0)}_{[j]}, ~\widetilde{\sigma}^{(0)}_{[j]}\in
\mathcal{A}_{[j]}(\varepsilon), \label{rela-0}
\end{equation}
and $\widetilde{\mathcal{L}}$ provides a self-transformation for
both $\mathcal{A}_{[0]}(\varepsilon)$ and
$\mathcal{A}_{[1]}(\varepsilon)$, i.e., $\forall \alpha\in
\mathcal{A}_{[0]}(\varepsilon)$, $\widetilde{\mathcal{L}}\alpha\in
\mathcal{A}_{[0]}(\varepsilon)$, and $\forall \beta\in
\mathcal{A}_{[1]}(\varepsilon)$, $\widetilde{\mathcal{L}}\beta\in
\mathcal{A}_{[1]}(\varepsilon)$. That means
\begin{equation}
\widetilde{K}^{(l)}_{[j]}, ~\widetilde{\sigma}^{(s)}_{[j]}\in
\mathcal{A}_{[j]}(\varepsilon). \label{rela-l-s}
\end{equation}

Noting that $\widetilde{u}_n=u_n|_{R_n=-\varepsilon
Q_n^*}=(Q_n,-\varepsilon Q_n^*)^T\in
\mathcal{A}_{[1]}(\varepsilon)$, we then define the following
isospectral IDNLS hierarchy
\begin{subequations}
\begin{equation}
i^{1-j}\widetilde{u}_{n,t_{m,j}}=\widetilde{K}^{(m)}_{[j]}=\widetilde{\mathcal{L}}^m\widetilde{K}^{(0)}_{[j]},~~
m=0,1,2,\cdots, \label{idnlse-hie-i}
\end{equation}
and non-isospectral IDNLS hierarchy
\begin{equation}
i^{1-j}\widetilde{u}_{n,t_{s,j}}=\widetilde{\sigma}^{(s)}_{[j]}=\widetilde{\mathcal{L}}^s
\widetilde{\sigma}^{(0)}_{[j]},~~ s=0,1,2,\cdots,
\label{idnlse-hie-n}
\end{equation}
\end{subequations}
where $j\in\{0,1\}$, $\widetilde{\mathcal{L}}$ is the recursion
operator, and we add subindexes $m,j$ and $s,j$ for $t$ as labels of
equations in their hierarchy. In fact,
the IDNLS equation \eqref{idnlse}
comes from \eqref{idnlse-hie-i} with $j=0,m=1$, i.e.,
\begin{equation}
\label{idnlse-vec} i\widetilde{u}_{n,t_{1,0}}= i\left(
\begin{array}{c} Q_{n}\\-\varepsilon Q^*_{n}
\end{array}\right)_{t_{1,0}}
=\widetilde{K}^{(1)}_{[0]} =\left(
\begin{array}{c} Q_{n+1}+Q_{n-1}-2Q_{n}+\varepsilon Q_{n}Q^{*}_{n}(Q_{n+1}+Q_{n-1})\\
\varepsilon Q_{n+1}^{*}+\varepsilon Q_{n-1}^{*}-2\varepsilon
Q^{*}_{n}+Q_{n}Q^{*}_{n}(Q^{*}_{n+1}+Q^{*}_{n-1})
\end{array}\right).
\end{equation}
In non-isospectral case, \eqref{idnlse-hie-n} with $j=0$ and $s=1$
reads
\begin{align}
i\left(
\begin{array}{r} Q_{n}\\-\varepsilon Q^*_{n}
\end{array}\right)_{t_{1,0}}=\widetilde{\sigma}^{(1)}_{[0]}
=&\frac{1}{2}\left(
\begin{array}{c} (1+\varepsilon Q_{n}Q_{n}^{*}) [(2n+3)Q_{n+1}+(2n-1)Q_{n-1}]-2(2n+1)Q_{n}\\
\varepsilon(1+\varepsilon Q_{n}Q_{n}^{*})
[(2n+3)Q^{*}_{n+1}+(2n-1)Q_{n-1}^{*}]-2\varepsilon(2n+1)Q^{*}_{n}
\end{array}\right)\notag\\
&+\varepsilon\left(
\begin{array}{c} Q_{n}\\\varepsilon Q_{n}^{*}\end{array}\right)(E-1)^{-1}(Q_{n+1}Q_{n}^{*}+Q_{n}Q_{n+1}^{*}),
\end{align}
of which the first row provides a non-isospectral IDNLS equation
\begin{align}
iQ_{n,t_{1,0}}=&\frac{1}{2}(1+\varepsilon Q_{n}Q_{n}^{*})
[(2n+3)Q_{n+1}+(2n-1)Q_{n-1}]-(2n+1)Q_{n}\nonumber\\
&+\varepsilon Q_{n}(E-1)^{-1}(Q_{n+1}Q_{n}^{*}+Q_{n}Q_{n+1}^{*}),
\label{idnlse-n}
\end{align}
which goes to a non-isospectral NLS equation in continuous limit
(see Sec.\ref{sec-cl}).

The flows $\{i^{j-1}\widetilde{K}^{(l)}_{[j]}\}$ and
$\{i^{j-1}\widetilde{\sigma}^{(s)}_{[j]}\}$ defined in
\eqref{idnlse-flows} are called isospectral and non-isospectral
IDNLS flows in vector form, respectively. We add multiplier $i^{j-1}$ so that they are always in the set $A_{[1]}(\varepsilon)$
to which $\widetilde{u}_n$ belongs. The first components of
$\{i^{j-1}\widetilde{K}^{(l)}_{[j]}\}$ and
$\{i^{j-1}\widetilde{\sigma}^{(s)}_{[j]}\}$, which we respectively
denote them by $\{i^{j-1}\widetilde{K}^{(l)}_{[j],1}\}$ and
$\{i^{j-1}\widetilde{\sigma}^{(s)}_{[j],1}\}$, are called
isospectral and non-isospectral IDNLS flows in scalar form. Thus the
scalar form of isospectral and non-isospectral IDNLS hierarchies can
be written as
\begin{subequations}\label{idnlse-hie-sca}
\begin{align}
i^{1-j}Q_{n,t_{m,j}}&=\widetilde{K}^{(m)}_{[j],1},~~~~ m=0,1,2,\cdots,
\label{idnlse-hie-i-sca}\\
i^{1-j}Q_{n,t_{s,j}}&=\widetilde{\sigma}^{(s)}_{[j],1},~~~~
s=0,1,2,\cdots. \label{idnlse-hie-n-sca}
\end{align}
\end{subequations}

\subsection{Algebraic structures of the INDLS flows}
\label{sec-3.2}

Making use of \eqref{alg-K-sigma} one can derive the algebraic
relations for the IDNLS flows.

\begin{theorem}
\label{Th-alg-flows} Suppose that $Q_n$ is the only independent
variable and the Gateaux derivative is defined w.r.t. $Q_n$. Then
the scalar isospectral and non-isospectral IDNLS flows
$\{i^{j-1}\widetilde{K}^{(m)}_{[j],1}\}$ and
$\{i^{j-1}\widetilde{\sigma}^{(s)}_{[j],1}\}$ form a Lie algebra
through the Lie product $\k \cdot, \cdot \j_{_{Q_n}}$ with the
following structure
\begin{subequations}
\begin{align}
\k i^{j-1}\widetilde{K}^{(m)}_{[j],1}, i^{k-1}\widetilde{K}^{(s)}_{[k],1} \j_{_{Q_n}} &=0,\\
\k -i\widetilde{K}^{(m)}_{[0],1}, -i\widetilde{\sigma}^{(s)}_{[0],1} \j_{_{Q_n}} &=-2m \widetilde{K}^{(m+s-1)}_{[1],1}, \\
\k -i\widetilde{K}^{(m)}_{[0],1}, \widetilde{\sigma}^{(s)}_{[1],1}
\j_{_{Q_n}}
 &=-\frac{1}{2}im (\widetilde{K}^{(m+s+1)}_{[0],1}+4\widetilde{K}^{(m+s)}_{[0],1}), \label{K0-s1-1}\\
\k \widetilde{K}^{(m)}_{[1],1},
i^{j-1}\widetilde{\sigma}^{(s)}_{[j],1} \j_{_{Q_n}}
 &=\frac{1}{2}i^{j-1}[(m+1) \widetilde{K}^{(m+s+1)}_{[j],1}+2(2m+1)\widetilde{K}^{(m+s)}_{[j],1}], \\
\k -i\widetilde{\sigma}^{(m)}_{[0],1}, -i\widetilde{\sigma}^{(s)}_{[0],1} \j_{_{Q_n}} &=-2(m-s)\widetilde{\sigma}^{(m+s-1)}_{[1],1}, \\
\k i^{j-1}\widetilde{\sigma}^{(m)}_{[j],1},
\widetilde{\sigma}^{(s)}_{[1],1} \j_{_{Q_n}}
&=\frac{1}{2}i^{j-1}[(m-s-1+j)\widetilde{\sigma}^{(m+s+1)}_{[j],1}+2(2m-2s-1+j)\widetilde{\sigma}^{(m+s)}_{[j],1}],
%\k \widetilde{\sigma}^{(m)}_{[1],1},
%\widetilde{\sigma}^{(s)}_{[1],1} \j_{_{Q_n}}
%&=2(m-s)(\widetilde{\sigma}^{(m+s+1)}_{[1],1}+4\widetilde{\sigma}^{(m+s)}_{[1],1}),
\end{align}
\label{K-s-Lie}
\end{subequations}
where $j,k\in\{0,1\}$, $m,s\geq 0$ and we set
$\widetilde{K}^{(-1)}_{[j],1}=\widetilde{\sigma}^{(-1)}_{[j],1}=0$
once they appear on the r.h.s. of \eqref{K-s-Lie}.
We note that hereafter by $\k \cdot, \cdot \j_{_{Q_n}}$ we denote the
product defined through the Gateaux derivative w.r.t. $Q_n$.
\end{theorem}

We prove the theorem by two steps. First, we derive algebraic
structures for the following vector flows
\begin{equation}
K^{(l)}_{[j]}=\mathcal{L}^l K^{(0)}_{[j]},~~
\sigma^{(l)}_{[j]}=\mathcal{L}^l \sigma^{(0)}_{[j]},~~ j\in\{0,1\},
~~l=0,1,\cdots, \label{dakns-flows}
\end{equation}
where $K^{(0)}_{[j]}$ and $\sigma^{(0)}_{[j]}$ are given in
\eqref{K-j-0} and \eqref{sigma-j-0}, respectively.
$\{K^{(l)}_{[j]}\}$ and $\{\sigma^{(l)}_{[j]}\}$ are called
(semi-)DAKNS flows (see Sec.\ref{sec-5.1}) \footnote{In continuous
limit one can find $K^{(0)}_{[0]}\sim (q,-r)^{T},~K^{(0)}_{[1]}\sim
(q_x,r_x)^{T}, ~\sigma^{(0)}_{[0]}\sim
(xq,-xr)^{T},~\sigma^{(0)}_{[1]}\sim (q+xq_x,r+xr_x)^{T}$ and
$\mathcal{L}\sim L^2_{AKNS}$ where $L_{AKNS}$ is the recursion
operator of the AKNS system.}. For these flows we have
\begin{lemma}
\label{Lem-alg-flows} ~ The flows $\{K^{(m)}_{[j]}\}$ and
$\{\sigma^{(s)}_{[j]}\}$ form a Lie algebra, denoted by $\mathcal{D}$,   through $\k \cdot,
\cdot \j$ with structure
\begin{subequations}
\begin{align}
\k K^{(m)}_{[j]}, K^{(s)}_{[k]} \j &=0,\\
\k {K}^{(m)}_{[0]}, {\sigma}^{(s)}_{[0]} \j &=2m {K}^{(m+s-1)}_{[1]}, \\
\k {K}^{(m)}_{[0]}, {\sigma}^{(s)}_{[1]} \j &=\frac{1}{2}m( {K}^{(m+s+1)}_{[0]}+4{K}^{(m+s)}_{[0]}), \label{K0-s1}\\
\k {K}^{(m)}_{[1]}, {\sigma}^{(s)}_{[j]} \j &=\frac{1}{2}[(m+1){K}^{(m+s+1)}_{[j]}+2(2m+1){K}^{(m+s)}_{[j]}], \\
\k {\sigma}^{(m)}_{[0]}, {\sigma}^{(s)}_{[0]} \j &=2(m-s){\sigma}^{(m+s-1)}_{[1]}, \\
\k {\sigma}^{(m)}_{[j]}, {\sigma}^{(s)}_{[1]} \j
&=\frac{1}{2}[(m-s-1+j){\sigma}^{(m+s+1)}_{[j]}+2(2m-2s-1+j){\sigma}^{(m+s)}_{[j]}],
\end{align}
\label{K-s-Lie-akns}
\end{subequations}
where $j,k\in\{0,1\}$, the Gateaux derivative is still defined w.r.t. $u_n$, $m,s\geq 0$ and
we set
$K^{(-1)}_{[j]}=\sigma^{(-1)}_{[j]}=0$
once they appear on the r.h.s. of \eqref{K-s-Lie-akns}.
\end{lemma}

\begin{proof}
We only prove \eqref{K0-s1}. The others can be proved similarly.
Noting that
\begin{equation}
\mathcal{L}^m=(L-2I+L^{-1})^m=\sum^{m}_{r=0} \sum^{m-r}_{j=0}  C^{r}_{m}C^{j}_{m-r} (-2)^{j}L^{m-2r-j},
\label{L-tilde-m}
\end{equation}
and by this we write ${K}^{(m)}_{[0]}$ and ${\sigma}^{(s)}_{[1]}$ as
\begin{align*}
{K}^{(m)}_{[0]}& = \sum^{m}_{r=0} \sum^{m-r}_{j=0} C^{r}_{m} C^{j}_{m-r} (-2)^{j}K^{(m-2r-j)},\\
{\sigma}^{(s)}_{[1]}&= \frac{1}{4} \sum^{s}_{h=0} \sum^{s-h}_{k=0}
C^{h}_{s} C^{k}_{s-h} (-2)^{k}
(\sigma^{(s-2h-k+1)}-\sigma^{(s-2h-k-1)}).
\end{align*}
Substituting them into $\k {K}^{(m)}_{[0]}, {\sigma}^{(s)}_{[1]} \j$
and making using of the Lie product relation \eqref{alg-K-sigma-b} yield
\begin{equation}
\k {K}^{(m)}_{[0]}, {\sigma}^{(s)}_{[1]}\j=A+B+C,
\end{equation}
where
\begin{align*}
A &=\frac{1}{2}m(L-L^{-1})\sum^{m}_{r=0} \sum^{m-r}_{j=0} C^{r}_{m}
C^{j}_{m-r} (-2)^{j}
  \sum^{s}_{h=0} \sum^{s-h}_{k=0} C^{h}_{s} C^{k}_{s-h} (-2)^{k} K^{(m+s-2r-2h-j-k)},\\
B &=-(L-L^{-1})\sum^{m}_{r=0}\sum^{m-r}_{j=0}r C^{r}_{m}
C^{j}_{m-r} (-2)^{j}
  \sum^{s}_{h=0}\sum^{s-h}_{k=0} C^{h}_{s}  C^{k}_{s-h} (-2)^{k} K^{(m+s-2r-2h-j-k)},\\
C &=-\frac{1}{2}(L-L^{-1})\sum^{m}_{r=0}  \sum^{m-r}_{j=0}j
C^{r}_{m}C^{j}_{m-r} (-2)^{j}
  \sum^{s}_{h=0} \sum^{s-h}_{k=0} C^{h}_{s} C^{k}_{s-h} (-2)^{k}  K^{(m+s-2r-2h-j-k)}.
\end{align*}
Next, still using \eqref{L-tilde-m}, the first term $A$ is nothing but
\begin{equation*}
A=\frac{1}{2}m (L-L^{-1})\mathcal{L}^m \mathcal{L}^s
K^{(0)}=\frac{1}{2}m (L-L^{-1}) K^{(m+s)}_{[0]}.
\end{equation*}
For the second term $B$ where the summation for $r$ essentially starts from $r=1$, again using \eqref{L-tilde-m} we have
\begin{equation*}
B=-(L-L^{-1})\mathcal{L}^s \sum^{m}_{r=1} r C^{r}_{m} (L-2I)^{m-r}
L^{-r}K^{(0)}.
\end{equation*}
It then follows from  the formula $r C^{r}_{m}=m C^{r-1}_{m-1}$ that
\begin{align*}
B& =- (L-L^{-1})\mathcal{L}^s \sum^{m-1}_{r=0} m C^{r}_{m-1} (L-2I)^{m-1-r} L^{-r-1}K^{(0)}\\
& =-m (L-L^{-1})L^{-1}\mathcal{L}^s \mathcal{L}^{m-1} K^{(0)}=-m
(L-L^{-1})L^{-1} K^{(m+s-1)}_{[0]}.
\end{align*}
Similarly, for the last term $C$, using
$$j C^{r}_{m}C^{j}_{m-r}=(m-r) C^{r}_{m} C^{j-1}_{m-r-1}=mC^{r}_{m-1} C^{j-1}_{m-r-1}$$
we have
\begin{equation*}
C=m (L-L^{-1}) K^{(m+s-1)}_{[0]}.
\end{equation*}
Then we have
\begin{align*}
\k {K}^{(m)}_{[0]}, {\sigma}^{(s)}_{[1]}\j &=A+B+C\\
&=\frac{1}{2}m(L-L^{-1})(\mathcal{L}-2L^{-1}+2I) K^{(m+s-1)}_{[0]}\\
&=\frac{1}{2}m(L-L^{-1})^2 K^{(m+s-1)}_{[0]}\\
&=\frac{1}{2}m(\mathcal{L}^2+4\mathcal{L}) K^{(m+s-1)}_{[0]}\\
&= \frac{1}{2}m (K^{(m+s+1)}_{[0]} + 4 K^{(m+s)}_{[0]}),
\end{align*}
which is just \eqref{K0-s1}. The other relations in
\eqref{K-s-Lie-akns} can be proved in a similar way.
\end{proof}
It is easy to find that
the Lie algebra $\mathcal{D}$ is
generated by the following elements
\begin{equation}
\{K^{(0)}_{[0]},~~ K^{(0)}_{[1]}~(or ~K^{(1)}_{[0]}),~~
\sigma^{(0)}_{[0]},~~ \sigma^{(0)}_{[1]} ~(or
~\sigma^{(1)}_{[0]}),~~ \sigma^{(1)}_{[1]}\}.
\end{equation}

The second step consists of a discussion for the consistency of the
reduction of \eqref{K-s-Lie-akns} under $R_n=-\varepsilon Q_n^*$.
Let us first consider 2-dimensional vector functions:
\begin{equation}
f(u_n)=(f_1,f_2)^T,~~~g(u_n)=(g_1,g_2)^T,~~~h(u_n)=(h_1,h_2)^T,
\end{equation}
which are related by
\begin{equation}
\k f,g\j=h. \label{fgh}
\end{equation}
We note that if $Q_n$ and $R_n$ are two independent variables and
there is no complex operation of $Q_n$ and $R_n$ in $f,g$, then the
linear relationship holds on the complex number field $\mathbb{C}$,
i.e.,
\begin{equation}
f(u_n)'[ag]=af(u_n)'[g], ~~\forall a\in \mathbb{C}.
\end{equation}
However, when $R_n=-\varepsilon Q_n^*$ and $Q_n$ is considered to be
the only one independent variable, in general the above linear
relationship does not hold any longer unless $a$ is
real.\footnote{For example, $f_1=Q_n^2+Q_n^*,~ g_1=Q_{n,x}$,
$f_1(Q_n)'[g_1]=2Q_nQ_{n,x}+Q_{n,x}^*$ but
$f_1(Q_n)'[ig_1]=i(2Q_nQ_{n,x}-Q_{n,x}^*)\neq if_1(Q_n)'[g_1]$.}

For a reasonable reduction for \eqref{K-s-Lie-akns} the problem we have
to conquer is the `consistency' of the product \eqref{fgh}:
\begin{itemize}
\item{First, a consistent reduction
requires two components are somehow related after reduction, for
example, $h_2=-\varepsilon h_1^*$.}
\item{
Second, when $Q_n$ becomes the only one independent variable in stead of
$(Q_n,R_n)$, a consistent reduction for the product
\eqref{fgh}$|_{R_n=-\varepsilon Q_n^*}$ should provide
\begin{equation}
\k f_1,g_1\j_{_{Q_n}}=f_1(Q_n)'[g_1]-g_1(Q_n)'[f_1]=h_1.
\label{fgh-1}
\end{equation}}
\end{itemize}
Such consistency for the product  \eqref{fgh} can be guaranteed by
taking $f(\widetilde{u}_n),g(\widetilde{u}_n)$ and $\widetilde{u}_n$ are in the same function set, i.e.,
\begin{equation}
f(\widetilde{u}_n),g(\widetilde{u}_n) \in A_{[1]}(\varepsilon),
\end{equation}
same as $\widetilde{u}_n$.

After the above discussion for the consistency of reduction, first,
we multiply $K^{(m)}_{[j]}$ and $\sigma^{(s)}_{[j]}$ on the l.h.s.
of \eqref{K-s-Lie-akns} by $i^{j-1}$ which just guarantees
$\{i^{j-1}\widetilde{K}^{(l)}_{[j]}\}$ and
$\{i^{j-1}\widetilde{\sigma}^{(s)}_{[j]}\}$ are in the set
$A_{[1]}(\varepsilon)$ to which $\widetilde{u}_n$ belongs. Then we
take the reduction $R_n=-\varepsilon Q_n^*$  and following
\eqref{fgh-1} we get the algebraic relations for those first
components, which are listed in Theorem \ref{Th-alg-flows}. Taking
\eqref{K0-s1} as an example, we first multiply ${K}^{(m)}_{[0]}$ by $-i$ and then rewrite \eqref{K0-s1} to
\begin{equation}
\k -i{K}^{(m)}_{[0]}, {\sigma}^{(s)}_{[1]} \j =-\frac{1}{2}im(
{K}^{(m+s+1)}_{[0]}+4{K}^{(m+s)}_{[0]}). \label{K0-s1+i}
\end{equation}
This is then ready for a consistent reduction and after taking
$R_n=-\varepsilon Q_n^*$ we get \eqref{K0-s1-1}.

\section{Symmetries}\label{sec-4}

\subsection{Symmetries for the IDNLS equation }

With the algebraic relations \eqref{K-s-Lie} in hand, we can
construct symmetries for the IDNLS equation \eqref{idnlse}, i.e.,
$iQ_{n,t_{1,0}}= \widetilde{K}^{(1)}_{[0],1}$. A scalar function
$\tau=\tau(Q_n)$ is a symmetry of \eqref{idnlse}, if
\begin{equation}
\tau_{t_{1,0}}=-i\widetilde{K}^{(1)}_{[0],1}(Q_n)'[\tau],
\label{4.1}
\end{equation}
which is, equivalently,
\begin{equation}
\frac{\tilde{\partial} \tau}{\tilde{\partial} t_{1,0}}=\K
-i\widetilde{K}^{(1)}_{[0],1}, \tau\J   _{_{Q_n}}, \label{def-sym}
\end{equation}
where    $\frac{\tilde{\partial}\tau}{\tilde{\partial} t_{1,0}}$
specially denotes the derivative of $\tau$ w.r.t. $t_{1,0}$ explicitly
included in  $\tau$ (cf. \cite{ZDJ-AL-2010}),
and the Gateaux derivative in \eqref{4.1} is defined w.r.t. $Q_n$.

From the algebraic structures in Theorem \ref{Th-alg-flows} and the
definition \eqref{def-sym} we have the following symmetries for the
IDNLS equation \eqref{idnlse}:   $K$-symmetries
$\{i^{j-1}\widetilde{K}^{(m)}_{[j],1}\}$ and $\tau$-symmetries
\begin{equation}
\tau^{(1,s)}_{[0,j]}=t_{1,0} \cdot \K
-i\widetilde{K}^{(1)}_{[0],1},
i^{j-1}\widetilde{\sigma}^{(s)}_{[j],1}\J   _{_{Q_n}}+i^{j-1}\widetilde{\sigma}^{(s)}_{[j],1},
~~s=0,1,\cdots, \label{tau-sym}
\end{equation}
i.e.,
\begin{subequations}
\begin{align}
\tau^{(1,s)}_{[0,0]}&=-2 t_{1,0} \widetilde{K}^{(s)}_{[1],1}-i \widetilde{\sigma}^{(s)}_{[0],1},\\
\tau^{(1,s)}_{[0,1]}&=-\frac{1}{2}i t_{1,0}(
\widetilde{K}^{(s+2)}_{[0],1}+4\widetilde{K}^{(s+1)}_{[0],1})+\widetilde{\sigma}^{(s)}_{[1],1}.
\end{align}
\label{tau-sym-exp}
\end{subequations}

The algebraic relations in \eqref{K-s-Lie} suggest an algebra  for the
symmetries of the IDNLS equation \eqref{idnlse}. This is concluded
by
\begin{theorem}
\label{Th-alg-sym} The isospectral IDNLS equation \eqref{idnlse} can
have two sets of symmetries,  $K$-symmetries
$\{i^{l-1}\widetilde{K}^{(m)}_{[l],1}\}$ and $\tau$-symmetries
$\tau^{(1,s)}_{[0,l]}$ given in \eqref{tau-sym-exp}, which form a
Lie algebra with structure
\begin{subequations}
\label{alg-sym}
\begin{align}
\k i^{z-1}\widetilde{K}^{(m)}_{[z],1}, i^{l-1}\widetilde{K}^{(s)}_{[l],1} \j_{_{Q_n}} &=0,\\
\k -i\widetilde{K}^{(m)}_{[0],1}, \tau^{(1,s)}_{[0,0]} \j_{_{Q_n}} &=-2m \widetilde{K}^{(m+s-1)}_{[1],1}, \\
\k -i\widetilde{K}^{(m)}_{[0],1}, \tau^{(1,s)}_{[0,1]} \j_{_{Q_n}}
 &=-\frac{1}{2}im( \widetilde{K}^{(m+s+1)}_{[0],1}+4\widetilde{K}^{(m+s)}_{[0],1}), \\
\k \widetilde{K}^{(m)}_{[1],1}, \tau^{(1,s)}_{[0,l]}  \j_{_{Q_n}}
 &=\frac{1}{2} i^{l-1}[(m+1)\widetilde{K}^{(m+s+1)}_{[l],1}+2(2m+1)\widetilde{K}^{(m+s)}_{[l],1}], \\
 \k \tau^{(1,m)}_{[0,0]} , \tau^{(1,s)}_{[0,0]}  \j_{_{Q_n}}
&=-2(m-s)\tau^{(1,m+s-1)}_{[0,1]},\\
\k \tau^{(1,m)}_{[0,l]} , \tau^{(1,s)}_{[0,1]}  \j_{_{Q_n}}
&=\frac{1}{2}[(m-s-1+l)\tau^{(1,m+s+1)}_{[0,l]}+2(2m-2s-1+l)\tau^{(1,m+s)}_{[0,l]}],
\end{align}
\end{subequations}
where $z,l\in\{0,1\}$, $m,s\geq 0$ and we set
$\widetilde{K}^{(-1)}_{[l],1}=\tau^{(1,-1)}_{[0,l]}=0$ once they
appear on the r.h.s. of \eqref{alg-sym}.
\end{theorem}

\subsection{Symmetries for the isospectral IDNLS hierarchy}

The algebraic structures in \eqref{K-s-Lie} also enable us to get
two sets of  symmetries of the isospectral IDNLS hierarchy
\eqref{idnlse-hie-i-sca}. For this we have the following theorem.

\begin{theorem}
\label{Th-alg-sym-hie} Each equation
$i^{1-j}Q_{n,t_{k,j}}=\widetilde{K}^{(k)}_{[j],1}$ in the
isospectral IDNLS hierarchy \eqref{idnlse-hie-i-sca} has two sets of
symmetries. When $j=0$ these symmetries are
\begin{subequations}\label{sym-d-IDNLS-0}
\begin{align}
K\hbox{-symmetries:~}& \{i^{l-1}\widetilde{K}^{(m)}_{[l],1}\},~~l\in\{0,1\},\\
\tau\hbox{-symmetries:~}& \tau^{(k,s)}_{[0,0]}=-2 k\,t_{k,0} \widetilde{K}^{(k+s-1)}_{[1],1}-i \widetilde{\sigma}^{(s)}_{[0],1},\label{tau-sym-d-IDNLS-0-0}\\
                        & \tau^{(k,s)}_{[0,1]}=-\frac{1}{2}i k\,t_{k,0}(
\widetilde{K}^{(k+s+1)}_{[0],1}+4\widetilde{K}^{(k+s)}_{[0],1})+\widetilde{\sigma}^{(s)}_{[1],1};\label{tau-sym-d-IDNLS-0-1}
\end{align}
\end{subequations}
and when $j=1$ the symmetries are
\begin{subequations}\label{sym-d-IDNLS-0}
\begin{align}
K\hbox{-symmetries:~}& \{i^{l-1}\widetilde{K}^{(m)}_{[l],1}\},~~l\in\{0,1\},\\
\tau\hbox{-symmetries:~}& \tau^{(k,s)}_{[1,0]}=-\frac{1}{2}i
(k+1)t_{k,1} \widetilde{K}^{(k+s+1)}_{[0],1}-i(2k+1)t_{k,1}
\widetilde{K}^{(k+s)}_{[0],1} -i\widetilde{\sigma}^{(s)}_{[0],1},\label{tau-sym-d-IDNLS-1-0}\\
                        & \tau^{(k,s)}_{[1,1]}=\frac{1}{2} (k+1)t_{k,1}
\widetilde{K}^{(k+s+1)}_{[1],1}+(2k+1)t_{k,1}
\widetilde{K}^{(k+s)}_{[1],1}
+\widetilde{\sigma}^{(s)}_{[1],1}.\label{tau-sym-d-IDNLS-1-1}
\end{align}
\end{subequations}
Symmetries for each equation can form a Lie algebra and structures are described as
\begin{subequations}
\label{alg-sym-hie}
\begin{align}
\k i^{z-1}\widetilde{K}^{(m)}_{[z],1}, i^{l-1}\widetilde{K}^{(s)}_{[l],1} \j_{_{Q_n}} &=0,\\
\k -i\widetilde{K}^{(m)}_{[0],1}, \tau^{(k,s)}_{[j,0]} \j_{_{Q_n}} &=-2m \widetilde{K}^{(m+s-1)}_{[1],1}, \\
\k -i\widetilde{K}^{(m)}_{[0],1}, \tau^{(k,s)}_{[j,1]} \j_{_{Q_n}}
 &=-\frac{1}{2}im( \widetilde{K}^{(m+s+1)}_{[0],1}+4\widetilde{K}^{(m+s)}_{[0],1}), \\
 \k \widetilde{K}^{(m)}_{[1],1}, \tau^{(k,s)}_{[j,l]}  \j_{_{Q_n}}
 &=\frac{1}{2}i^{l-1}[(m+1)\widetilde{K}^{(m+s+1)}_{[l],1}+2(2m+1)\widetilde{K}^{(m+s)}_{[l],1}], \\
 \k \tau^{(k,m)}_{[j,0]} , \tau^{(k,s)}_{[j,0]}  \j_{_{Q_n}}
&=-2(m-s) \tau^{(k,m+s-1)}_{[j,1]},\\
\k \tau^{(k,m)}_{[j,l]} , \tau^{(k,s)}_{[j,1]}  \j_{_{Q_n}}
&=\frac{1}{2}[(m-s-1+l)\tau^{(k,m+s+1)}_{[j,l]}+2(2m-2s-1+l)\tau^{(k,m+s)}_{[j,l]}],
\end{align}
\end{subequations}
where $j,z,l\in\{0,1\}$, $k,m,s\geq 0$ and we set
$\widetilde{K}^{(-1)}_{[l],1}=\tau^{(k,-1)}_{[j,l]}=0$ once they
appear on the r.h.s. of \eqref{alg-sym-hie}. Especially, when $j=0$,
$k=1$ the above results reduce to Theorem \ref{Th-alg-sym}.
\end{theorem}

\subsection{Relations between flows and the recursion operator $\mathcal{L}$ }

\begin{theorem}
\label{Th-recur-op} The flows
$\{{K}^{(m)}_{[j]}\}$ and $\{{\sigma}^{(m)}_{[j]}\}$ and their
recursion operator $\mathcal{L}$ satisfy
\begin{subequations}
\begin{align}
\mathcal{L}^{'}[{K}^{(m)}_{[j]}]-[{K}^{(m)'}_{[j]},\mathcal{L}]=0,~~~j\in\{0,1\},\\
\mathcal{L}^{'}[{\sigma}^{(m)}_{[0]}]-[{\sigma}^{(m)'}_{[0]},\mathcal{L}]-\mathcal{L}^{m}(L-L^{-1})=0,\\
\mathcal{L}^{'}[{\sigma}^{(m)}_{[1]}]-[{\sigma}^{(m)'}_{[1]},\mathcal{L}]-\frac{1}{2}\mathcal{L}^{m+2}-2\mathcal{L}^{m+1}=0,\label{L-1}
\end{align}
\label{L-flows}
\end{subequations}
where $m=0,1,2,\cdots$.
\end{theorem}

\begin{proof}
 We only prove \eqref{L-1}. The other two can be proved similarly.
We start from the relation
\begin{subequations}
\label{L-flows-old}
\begin{align}
 L'[\sigma^{(m)}]-[\sigma^{(m)'}, L]-2L^{m+1}&=0,\label{L-flows-old-a}\\
 (L^{-1})'[\sigma^{(m)}]-[\sigma^{(m)'}, L^{-1}]+2L^{m-1}&=0,\label{L-flows-old-b}
\end{align}
\end{subequations}
in which \eqref{L-flows-old-a} was given in Ref.\cite{ZDJ-06-PLA}
and \eqref{L-flows-old-b} can be proved similarly but we here skip
the proof. We can express
$\mathcal{L}^{'}[{\sigma}^{(m)}_{[1]}]-[{\sigma}^{(m)'}_{[1]},\mathcal{L}]$
in terms of $L, L^{-1}$ and ${\sigma}^{(s)}$ and then making use of
the above relation we find
\begin{align*}
& \mathcal{L}^{'}[{\sigma}^{(m)}_{[1]}]-[{\sigma}^{(m)'}_{[1]},\mathcal{L}]\\
=&\frac{1}{2} \sum^{m}_{h=0} \sum^{m-h}_{k=0} C^{h}_{m} C^{k}_{m-h}(-2)^{k}
(L^{m-2h-k+2}-L^{m-2h-k}-L^{m-2h-k}+(L^{-1})^{m-2h-k-2})\\
=& \frac{1}{2}
(L^{2}\mathcal{L}^{m}-\mathcal{L}^{m}-\mathcal{L}^{m}+L^{-2}\mathcal{L}^{m})\\
=&\frac{1}{2}\mathcal{L}^{m+2}+2\mathcal{L}^{m+1}.
\end{align*}
Thus we complete the proof.
\end{proof}

We note that these relations \eqref{L-flows} can be employed to prove the Lemma \ref{Lem-alg-flows}
if we use inductive approach (cf. \cite{Li-YS-86,Tian-book-90} for continuous cases).

%\newpage
\section{Continuous limit}
\label{sec-cl}

\subsection{DAKNS flows}\label{sec-5.1}

In Sec.\ref{sec-3.2} we introduced flows $\{K^{(l)}_{[j]}\}$ and
$\{\sigma^{(l)}_{[j]}\}$ given by  \eqref{dakns-flows}, which were referred
to as DAKNS flows. In fact, in continuous limit these flows just go
to the continuous AKNS isospectral and non-isospectral flows.

Let us consider the following limit (cf.\cite{Ablowitz-04-book}):
\begin{itemize}
\item{replacing $Q_n$ and $R_n$ with $hq_n$ and $hr_n$, where $h$ is the real step parameter (the lattice spacing),}
\item{$n\to \infty,~~h\to 0$ such that $nh$ finite,}
\item{introducing continuous variable $x=x_0+nh$,
then for a scalar function, for example, $q_n$, one has $q_{n+j}=q(x+jh)$.
For convenience we take $x_0=0$.}
\end{itemize}

Following the above limit procedure, one can find that
\begin{equation}
K^{(0)}_{[0]}\to \left(\begin{array}{c}q\\ -r\end{array}\right)
\label{akns-1}
\end{equation}
and the leading term is of $O(h)$;
\begin{equation}
K^{(0)}_{[1]}\to \left(\begin{array}{c}q\\ r\end{array}\right)_x
\label{akns-2}
\end{equation}
and the leading term is of $O(h^2)$. Besides, for any given scalar
functions $f_{1,n}$ and $f_{2,n}$, applying the above continuous
limit on the operator $\mathcal{L}$ \eqref{L-AKNS} and defining
integrate operator $\partial^{-1}_x\sim h(E-1)^{-1}$, one can find
that the leading term is of $O(h^2)$, which gives
\begin{align}
\mathcal{L} \left(\begin{array}{c}f_{1,n}\\ f_{2,n}\end{array}\right)
\to &  \Bigl(\partial^2_x-4qr
-2\left(\begin{array}{c}q_x\\ r_x\end{array}\right)\partial^{-1}_x(r,q)
-2\left(\begin{array}{c}q\\ -r\end{array}\right)\partial^{-1}_x(-r_x,q_x)\Bigr)
\left(\begin{array}{c} f_1(x)\\f_2(x)
\end{array}\right)
\nonumber\\
&= L_{AKNS}^{2}\left(\begin{array}{c} f_1(x)\\f_2(x)
\end{array}\right),
\label{cl-L}
\end{align}
where $L_{AKNS}$ is the well known recursion operator of the AKNS system, defined as (cf.\cite{AKNS,Li-YS-86})
\begin{equation}
L_{AKNS}=-\sigma_3\partial_x+2\sigma_3\left(\begin{array}{c}q\\
r\end{array}\right)\partial^{-1}_x(r,q),
\end{equation}
in which $\sigma_3$ is the Pauli matrix $\left(\begin{smallmatrix}-1&0\\0&1\end{smallmatrix}\right)$.
This result means in continuous limit $\mathcal{L}$ goes to the square of the AKNS recursion operator.
Then, noting that \eqref{akns-1} and \eqref{akns-2} are nothing but the first two flows in the AKNS hierarchy (cf.\cite{Chen-book,Li-YS-86}),
now it is clear that in continuous limit
\begin{subequations}
\begin{align}
K^{(m)}_{[0]}&\to K^{(2m)}_{AKNS},~~~~\mathrm{leading~ term} ~O(h^{2m+1}),\\
K^{(m)}_{[1]}&\to K^{(2m+1)}_{AKNS},~~~\mathrm{leading~ term} ~O(h^{2m+2}),
\end{align}
\label{akns-order}
\end{subequations}
where $\{K^{(s)}_{AKNS}\}$ are  the hierarchy of the AKNS isospectral flows.

After similar discussions, for the non-isospectral flows $\{\sigma^{(m)}_{[j]}\}$ we have
\begin{subequations}
\begin{align}
\sigma^{(0)}_{[0]}&\to \left(\begin{array}{c}xq\\ -xr\end{array}\right),~~~\mathrm{leading~ term} ~O(1),\\
\sigma^{(0)}_{[1]}&\to \left(\begin{array}{c}xq\\ xr\end{array}\right)_x,~~~~\mathrm{leading~ term} ~O(h),
\end{align}
\end{subequations}
which are the first two AKNS non-isospectral flows (cf.\cite{ZDJ-PA-IST}), and
\begin{subequations}
\begin{align}
\sigma^{(m)}_{[0]}&\to \sigma^{(2m)}_{AKNS},~~~~\mathrm{leading~ term} ~O(h^{2m}),\\
\sigma^{(m)}_{[1]}&\to \sigma^{(2m+1)}_{AKNS},~~~\mathrm{leading~ term} ~O(h^{2m+1}),
\end{align}
\label{akns-order-sig}
\end{subequations}
where $\{\sigma^{(s)}_{AKNS}\}$ are the hierarchy of the AKNS non-isospectral flows.

\subsection{Isospectral AKNS hierarchy and NLS hierarchy}

Based on the above discussion on continuous limit, we can define the
DAKNS hierarchy as
\begin{equation}
u_{n,t_{m,j}}=K^{(m)}_{[j]},~~~j\in \{0,1\}, ~~m=0,1,\cdots,
\label{akns-d-hie}
\end{equation}
which is an isospectral evolution equation hierarchy. Consider
continuous limit of the above hierarchy. The dominated terms on both
sides should have same order in terms of $h$. Since we have replaced
$u_n$ by $h\cdot(q,r)^T$, we still need to replace $t_{m,j}$ by
$t_{2m+j}\cdot h^{-(2m+j)}$, i.e.,
\begin{equation}
t_{m,j} \to t_{2m+j}\cdot h^{-(2m+j)},
\label{t-order}
\end{equation}
so that the left side of \eqref{akns-d-hie} is  $h^{2m+1+j}\cdot
\mathcal{U}_{t_{2m+j}}$, i.e., of $O(h^{2m+1+j})$, where
$\mathcal{U}=(q,r)^T$. Thus in continuous limit the DAKNS hierarchy
\eqref{akns-d-hie} goes to the ANKS isospectral evolution equation
hierarchy
\begin{equation}
\mathcal{U}_{t_{2m+j}}=K^{(2m+j)}_{AKNS},~~~j\in \{0,1\}, ~~m=0,1,\cdots.
\label{akns-c-hie}
\end{equation}

Then we define
\begin{equation}
\widetilde{K}^{(2m+j)}_{AKNS}=K^{(2m+j)}_{AKNS}|_{r=-\varepsilon q^{*}},~~~j\in \{0,1\}, ~~m=0,1,\cdots.
\end{equation}
Its first element $\widetilde{K}^{(2m+j)}_{AKNS,1}$ generates the NLS hierarchy
\begin{equation}
i^{1-j}q_{t_{2m+j}}=\widetilde{K}^{(2m+j)}_{AKNS,1},~~~j\in \{0,1\}, ~~m=0,1,\cdots,
\label{nls-hie}
\end{equation}
which is just the continuous limit of the IDNLS hierarchy \eqref{idnlse-hie-i-sca}.

\subsection{Symmetries}

Using the algebra $\mathcal{D}$ defined by \eqref{K-s-Lie-akns} one can construct two sets of symmetries for any equation
\begin{equation}
u_{n,t_{l,j}}=K^{(l)}_{[j]}
\label{akns-d-l}
\end{equation}
in the DAKNS hierarchy \eqref{akns-d-hie}. When $j=0$ these
symmetries are
\begin{subequations}\label{sym-d-akns-0}
\begin{align}
K\hbox{-symmetries:~}& K^{(m)}_{[k]},~~k\in\{0,1\},\\
\tau\hbox{-symmetries:~}& \mathcal{T}^{(l,s)}_{[0,0]}=2lt_{l,0}K_{[1]}^{(l+s-1)}+\sigma_{[0]}^{(s)},\label{tau-sym-d-akns-0-0}\\
                        & \mathcal{T}^{(l,s)}_{[0,1]}=\frac{1}{2}lt_{l,0}K_{[0]}^{(l+s+1)}+2lt_{l,0}K_{[0]}^{(l+s)}+\sigma_{[1]}^{(s)};\label{tau-sym-d-akns-0-1}
\end{align}
\end{subequations}
and when $j=1$ the symmetries are
\begin{subequations} \label{sym-d-akns-1}
\begin{align}
K\hbox{-symmetries:~}& K^{(m)}_{[k]},~~k\in\{0,1\},\\
\tau\hbox{-symmetries:~}&\mathcal{T}^{(l,s)}_{[1,0]}=\frac{1}{2}
(l+1)t_{l,1} K^{(l+s+1)}_{[0]}+(2l+1)t_{l,1} K^{(l+s)}_{[0]}
+\sigma^{(s)}_{[0]},\label{tau-sym-d-akns-1-0}\\
&\mathcal{T}^{(l,s)}_{[1,1]}=\frac{1}{2} (l+1)t_{l,1}
K^{(k+s+1)}_{[1]}+(2l+1)t_{l,1} K^{(l+s)}_{[1]}
+\sigma^{(s)}_{[1]}.\label{tau-sym-d-akns-1-1}
\end{align}
\end{subequations}
The symmetries for \eqref{akns-d-l} form a Lie algebra with structure
\begin{subequations}
\begin{align}
\k K^{(m)}_{[z]}, K^{(s)}_{[k]} \j &=0,\\
\k {K}^{(m)}_{[0]}, {\mathcal{T}}^{(l,s)}_{[j,0]} \j &=2m {K}^{(m+s-1)}_{[1]}, \\
\k {K}^{(m)}_{[0]}, {\mathcal{T}}^{(l,s)}_{[j,1]} \j &=\frac{1}{2}m( {K}^{(m+s+1)}_{[0]}+4{K}^{(m+s)}_{[0]}), \\
\k K^{(m)}_{[1]}, {\mathcal{T}}^{(l,s)}_{[j,k]} \j
 &=\frac{1}{2}(m+1)K^{(m+s+1)}_{[k]}+(2m+1)K^{(m+s)}_{[k]}, \\
 \k {\mathcal{T}}^{(l,m)}_{[j,0]} , {\mathcal{T}}^{(l,s)}_{[j,0]}  \j
&=2(m-s) {\mathcal{T}}^{(l,m+s-1)}_{[j,1]},\\
\k {\mathcal{T}}^{(l,m)}_{[j,k]} , {\mathcal{T}}^{(l,s)}_{[j,1]} \j
&=\frac{1}{2}(m-s-1+k){\mathcal{T}}^{(l,m+s+1)}_{[j,k]}+(2m-2s-1+k){\mathcal{T}}^{(l,m+s)}_{[j,k]},
\end{align}
\label{sym-alg-d-akns}
\end{subequations}
where $j,k,z\in\{0,1\}$, $l,m,s\geq 0$ and we set
$K^{(-1)}_{[z]}={\mathcal{T}}^{(l,-1)}_{[j,k]}=0$ once they appear
on the r.h.s. of \eqref{sym-alg-d-akns}.

For the continuous flows $\{K^{(m)}_{AKNS}\}$  and $\{\sigma^{(s)}_{AKNS}\}$, it has been known that
(cf.\cite{Li-YS-86,Tu-GZ-88,CDY-91,CDY-96}) they form an algebra as well.
We denote this algebra by $\mathcal{C}$. Its structure is
\begin{subequations}
\label{alg-K-sigma-akns}
\begin{align}
\k K^{(m)}_{AKNS}, K^{(s)}_{AKNS} \j &=0,\label{alg-K-sigma-a1}
\\
\k K^{(m)}_{AKNS}, \sigma^{(s)}_{AKNS} \j & =m K^{(m+s-1)}_{AKNS}, \label{alg-K-sigma-b1} \\
\k \sigma^{(m)}_{AKNS}, \sigma^{(s)}_{AKNS} \j
&=(m-s)\sigma^{(m+s-1)}_{AKNS}, \label{alg-K-sigma-c1}
\end{align}
\end{subequations}
where $m,s\geq 0$ and we set
$K^{(-1)}_{AKNS}=\sigma^{(-1)}_{AKNS}=0$ once they appear on the
r.h.s. of \eqref{alg-K-sigma-akns}. This algebra can be generated by
\begin{equation}
\{K^{(1)}_{AKNS}, ~\sigma^{(0)}_{AKNS}, ~\sigma^{(3)}_{AKNS}\}.
\end{equation}

From $\mathcal{C}$ one can have two sets of symmetries for each AKNS equation
\begin{equation}
\mathcal{U}_{t_{l}}=K^{(l)}_{AKNS}
\label{akns-c-l}
\end{equation}
in the hierarchy \eqref{akns-c-hie}, the symmetries are
\begin{subequations}
\begin{align}
K\hbox{-symmetries:~}& K^{(m)}_{AKNS},\\
\tau\hbox{-symmetries:~}& \mathcal{T}^{(l,s)}_{AKNS}=lt_{l}
K^{(l+s-1)}_{AKNS}+\sigma^{(s)}_{AKNS},\label{tau-sym-c-akns}
\end{align}
\label{sym-c-akns}
\end{subequations}
which form a Lie algebra with (cf.\cite{Li-YS-86,Tu-GZ-88,CDY-91,CDY-96})
\begin{subequations}
\begin{align}
\k K^{(m)}_{AKNS}, K^{(s)}_{AKNS} \j &=0,\\
\k {K}^{(m)}_{AKNS}, {\mathcal{T}}^{(l,s)}_{AKNS} \j &=m
K^{(m+s-1)}_{AKNS}, \\
\k {\mathcal{T}}^{(l,m)}_{AKNS}, {\mathcal{T}}^{(l,s)}_{AKNS} \j
&=(m-s){\mathcal{T}}^{(l,m+s-1)}_{AKNS},
\end{align}
\label{sym-alg-c-akns}
\end{subequations}
where $l,m,s\geq 0$ and we set
$K^{(-1)}_{AKNS}={\mathcal{T}}^{(l,-1)}_{AKNS}=0$ once they appear
on the r.h.s. of \eqref{sym-alg-c-akns}.

Staring from the relations \eqref{alg-K-sigma-akns} and employing similar discussions as we have done
for the IDNLS hierarchy in Sec.\ref{sec-4}, one can have two sets of symmetries for the $l$th equation in the NLS hierarchy \eqref{nls-hie}:
\begin{equation}
 q_{t_l}=\mu_l\widetilde{K}^{(l)}_{AKNS,1},~~~\mu_l=\left\{\begin{array}{ll}-i,& l~\mathrm{is~even},\\1,&l~\mathrm{is~odd}.\end{array}\right.
\label{nls-c-l}
\end{equation}
The symmetries are (cf.\cite{Li-YS-86})
\begin{subequations}
\begin{align}
K\hbox{-symmetries:~}& \mu_m\widetilde{K}^{(m)}_{AKNS,1},\\
\tau\hbox{-symmetries:~}& \widetilde{\mathcal{T}}^{(l,s)}=\mu_l
\mu_s l t_{l} \widetilde{K}^{(l+s-1)}_{AKNS,1}+\mu_s
\widetilde{\sigma}^{(s)}_{AKNS,1},
\end{align}
\label{sym-c-nls}
\end{subequations}
which compose  a Lie algebra with structure
\begin{subequations}
\begin{align}
\k \mu_m\widetilde{K}^{(m)}_{AKNS,1}, \mu_s\widetilde{K}^{(s)}_{AKNS,1} \j_{q} &=0,\\
\k \mu_m\widetilde{K}^{(m)}_{AKNS,1},
{\widetilde{\mathcal{T}}}^{(l,s)} \j_{q} &=m\mu_m \mu_s \widetilde{K}^{(m+s-1)}_{AKNS,1}, \\
\k {\widetilde{\mathcal{T}}}^{(l,m)},
{\widetilde{\mathcal{T}}}^{(l,s)} \j_{q} &=(m-s)\frac{\mu_m \mu_s
}{\mu_{m+s-1}}{\widetilde{\mathcal{T}}}^{(l,m+s-1)} ,
\end{align}
\label{sym-alg-c-nls}
\end{subequations}
where $l,m,s\geq 0$ and we set
$\widetilde{K}^{(-1)}_{AKNS,1}={\widetilde{\mathcal{T}}}^{(l,-1)}=0$
once they appear on the r.h.s. of \eqref{sym-alg-c-nls}. In the
product $\k \cdot,\cdot\j_q$ the Gateaux derivative is defined
w.r.t. $q$. When $l=1$, they are reduced the symmetries and algebra
for the NLS equation.

The symmetries \eqref{sym-d-akns-0}, \eqref{sym-d-akns-1} and
\eqref{sym-c-akns} are related together by continuous limit. Same
relations hold for the symmetries in Theorem \ref{Th-alg-sym-hie}
and \eqref{sym-c-nls}. We skip the detailed discussions for these
connections.

\section{Algebra deformations and understanding}

We have presented symmetries and their algebras for the continuous AKNS and NLS hierarchies.
Comparing them with those for the discrete cases, one can find not only the form of symmetries
but also the structures of algebras are different.
Since the algebras $\mathcal{D}$ and $\mathcal{C}$ (see \eqref{K-s-Lie-akns} and \eqref{alg-K-sigma-akns})
play key roles for generating symmetries, let us focus on
$\mathcal{D}$ and $\mathcal{C}$ and see the difference between them in the light of the correspondence \eqref{akns-order} and \eqref{akns-order-sig}.
Several of these deformations from $\mathcal{D}$ and $\mathcal{C}$ are listed in the following:
\begin{itemize}
\item{different structures and different generators;}
\item{$\{K^{(0)}_{[0]},K^{(0)}_{[1]},K^{(1)}_{[0]},\sigma^{(0)}_{[0]}\},
\{\sigma^{(0)}_{[0]},\sigma^{(0)}_{[1]},\sigma^{(1)}_{[0]}\}$ and
$\{K^{(0)}_{AKNS},K^{(1)}_{AKNS},K^{(2)}_{AKNS},\sigma^{(0)}_{AKNS}\},$\\
$\{\sigma^{(0)}_{AKNS},\sigma^{(1)}_{AKNS},\sigma^{(2)}_{AKNS}\}$
are subalgebras of $\mathcal{D}$ and $\mathcal{C}$ respectively but
with different structures;}
\item{
$\{K^{(0)}_{AKNS},K^{(1)}_{AKNS},\sigma^{(0)}_{AKNS},\sigma^{(1)}_{AKNS}\}$
is a subalgebra of  $\mathcal{C}$, but for $\mathcal{D}$ we do not find any similar
subalgebras (containing at least two non-isospectral flows and one isospectral flow).}
\end{itemize}

To understand these deformations, we introduce \textit{degree} for flows $\{{K}^{(m)}_{[j]}\}$ and $\{{\sigma}^{(m)}_{[j]}\}$,
variable $t_{m,j}$ and the recursion operator $\mathcal{L}$.
For a function $f(n,h,t)$ (or an operator) by deg$\,f$ we mean the order of $h$ of the denominate term (or leading term) in
continuous limit. So we can define
\begin{subequations}
\begin{align}
\mathrm{deg}\,{K}^{(m)}_{[j]}&= 2m+1+j,\\
\mathrm{deg}\,{\sigma}^{(m)}_{[j]}&= 2m+j,\\
\mathrm{deg}\,\mathcal{L}&= 2,\\
\mathrm{deg}\,t_{m,j}&= -(2m+j),\label{deg-t}\\
\mathrm{deg}\,u_n &= 1,\label{deg-u}
\end{align}
\label{deg}
\end{subequations}
where \eqref{deg-t} is from \eqref{t-order}.
Thus, after taking continuous limit only the terms with the lowest degree are left while others disappear.
We also note that due to \eqref{deg-u} and the definition of Gateaux derivative
\begin{equation}
\mathrm{deg}\,f(u_n)'[g(u_n)]=\mathrm{deg}\,\k f(u_n),g(u_n)\j=\mathrm{deg}\, f(u_n)+ \mathrm{deg}\,g(u_n)-\mathrm{deg}\,u_n.
\end{equation}

Now let us take \eqref{K0-s1} and \eqref{alg-K-sigma-b1} as an example to see the role that the degrees play
in continuous limit and understanding those deformations. \eqref{K0-s1} reads
\begin{equation}
\k {K}^{(m)}_{[0]}, {\sigma}^{(s)}_{[1]} \j =\frac{1}{2}m(
{K}^{(m+s+1)}_{[0]}+4{K}^{(m+s)}_{[0]}),\label{K0-s1-new}
\end{equation}
in which
\begin{equation}
\mathrm{deg}\,\k {K}^{(m)}_{[0]}, {\sigma}^{(s)}_{[1]} \j
=2(m+s)+1,~~ \mathrm{deg}\,{K}^{(m+s+1)}_{[0]}=2(m+s)+3,~~
\mathrm{deg}\,{K}^{(m+s)}_{[0]}=2(m+s)+1.
\end{equation}
With the help of \textit{degree} and noting that the correspondence
\eqref{akns-order} and \eqref{akns-order-sig}, after taking
continuous limit, only those terms with the lowest degree are left
and consequently \eqref{K0-s1-new} goes to
\begin{equation*}
\k {K}^{(2m)}_{AKNS}, {\sigma}^{(2s+1)}_{AKNS} \j
=2m{K}^{(2(m+s))}_{AKNS},
\end{equation*}
which belongs to relation \eqref{alg-K-sigma-b1}. Similar to this example we
can examine degrees of other formulas in \eqref{K-s-Lie-akns} and in
this way we can explain why the algebra $\mathcal{D}$ goes to
$\mathcal{C}$ in continuous limit although they have different
structures.

Next we turn to $\tau$-symmetries. Some of them contain different number of
terms in discrete case and continuous case. To explain the difference we consider
\eqref{tau-sym-d-akns-0-0}, \eqref{tau-sym-d-akns-0-1},
\eqref{tau-sym-d-akns-1-0}, \eqref{tau-sym-d-akns-1-1}  and
\eqref{tau-sym-c-akns} as an example. In \eqref{tau-sym-d-akns-0-0}
\begin{equation*}
\mathrm{deg}\,(t_{l,0}{K}^{(l+s-1)}_{[1]})= -2l+[2(l+s-1)+1+1]=2s,~~
\mathrm{deg}\,{\sigma}^{(s)}_{[0]}= 2s,
\end{equation*}
which are same. Thus in continuous limit \eqref{tau-sym-d-akns-0-0} yields
\begin{equation*}
\tau^{(2l,2s)}_{AKNS}=2lt_{2l} K^{(2(l+s)-1)}_{AKNS}+\sigma^{(2s)}_{AKNS}.
\end{equation*}
In \eqref{tau-sym-d-akns-0-1} the degrees are
\begin{equation*}
\mathrm{deg}\,(t_{l,0}{K}^{(l+s+1)}_{[0]})= 2s+3,~~
\mathrm{deg}\,(t_{l,0}{K}^{(l+s)}_{[0]})=
\mathrm{deg}\,{\sigma}^{(s)}_{[1]}= 2s+1.
\end{equation*}
Thus in continuous limit the term $t_{l,0}{K}^{(l+s+1)}_{[0]}$ will
disappear due to higher degree and then \eqref{tau-sym-d-akns-0-1} goes
to
\begin{equation*}
\tau^{(2l,2s+1)}_{AKNS}=2lt_{2l}
K^{(2(l+s))}_{AKNS}+\sigma^{(2s+1)}_{AKNS}.
\end{equation*}
Similarly, \eqref{tau-sym-d-akns-1-0} goes to
\begin{equation*}
\tau^{(2l+1,2s)}_{AKNS}=(2l+1)t_{2l+1}
K^{(2(l+s))}_{AKNS}+\sigma^{(2s)}_{AKNS},
\end{equation*}
and \eqref{tau-sym-d-akns-1-1} goes to
\begin{equation*}
\tau^{(2l+1,2s+1)}_{AKNS}=(2l+1)t_{2l+1}
K^{(2(l+s)+1)}_{AKNS}+\sigma^{(2s+1)}_{AKNS}.
\end{equation*}
The above four continuous limit results just compose the $\tau$-symmetry
\eqref{tau-sym-c-akns}.

Now by means of \textit{degree} we can understand the deformation of algebras and symmetries appearing in continuous limit.
Finally, let us look at the relation \eqref{L-flows}.
In continuous limit we have
\begin{subequations}
\begin{align}
(L^2_{AKNS})^{'}[{K}^{(m)}_{AKNS}]-[{K}^{(m)'}_{AKNS},L^2_{AKNS}]&=0,\\
(L^2_{AKNS})^{'}[{\sigma}^{(m)}_{AKNS}]-[{\sigma}^{(m)'}_{AKNS},L^2_{AKNS}]-2L^{m+1}_{AKNS}&=0.
\end{align}
\label{L-flows-cl}
\end{subequations}
This is consistent with the result for the AKNS recursion operator and flows:
\begin{subequations}
\begin{align}
L^{'}_{AKNS}[{K}^{(m)}_{AKNS}]-[{K}^{(m)'}_{AKNS},L_{AKNS}]&=0,\\
L^{'}_{AKNS}[{\sigma}^{(m)}_{AKNS}]-[{\sigma}^{(m)'}_{AKNS},L_{AKNS}]-L_{AKNS}^{m}&=0,
\end{align}
\label{L-flows-akns}
\end{subequations}
where we have made used of the result
\begin{equation}
\frac{1}{2}(L-L^{-1}) \to  L_{AKNS},~~~ \mathrm{leading ~term:}~ O(h)
\end{equation}
in continuous limit, which can be seen through a procedure similar
to \eqref{cl-L}. Applying $L_{AKNS}$ to \eqref{L-flows-akns} and
using the Leibniz rule for two operators $F$ and $G$: $(FG)'[h]=F'[h]G+F\,G'[h]$, one can
derive \eqref{L-flows-cl} from \eqref{L-flows-akns}. However, one
may wonder that now that in continuous limit we have the operator
$\frac{1}{2}(L-L^{-1}) \to L_{AKNS}$ but $\mathcal{L} \to
L^2_{AKNS}$, why we use $\mathcal{L}$ in stead of
$\frac{1}{2}(L-L^{-1})$ to generate DAKNS hierarchy? In fact, it is
true that different discrete flows can go to the same in continuous
limit. Applying $\frac{1}{2}(L-L^{-1})$ on $K^{(0)}_{[0]}$ twice and
taking $Q_n=-\varepsilon R_n^*$ we have a second IDNLS equation,
\begin{align}
iQ_{n,t_2}=&\frac{1}{4}(1+\varepsilon
Q_{n}Q_{n}^{*})[Q_{n+2}(1+\varepsilon
Q_{n+1}Q_{n+1}^{*})+\varepsilon Q_{n}Q_{n+1}Q_{n-1}^{*}+\varepsilon
Q_{n+1}^{2}Q_{n}^{*}\notag\\
&+Q_{n-2}(1+\varepsilon Q_{n-1}Q_{n-1}^{*})+\varepsilon
Q_{n-1}^{2}Q_{n}^{*}+\varepsilon
Q_{n-1}Q_{n}Q_{n+1}^{*}]-\frac{1}{2}Q_{n}.
\end{align}
It looks more complicated than the IDNLS equation \eqref{idnlse} but
it does go to the continuous NLS equation. Therefore we prefer to
$\mathcal{L}$ although it corresponds to $L^2_{AKNS}$.

\section{Conclusions}

One of the main results of the paper is that we got infinitely many
symmetries for the IDNLS equation \eqref{idnlse} and the IDNLS
hierarchy \eqref{idnlse-hie-i-sca}. This was done through
constructing the recursion operator $\widetilde{\mathcal{L}}$,
isospectral and non-isospectral IDNLS flows in scalar form and their
algebraic structures \eqref{K-s-Lie}.

A second result is on the DAKNS flows. These flows are generated by
the basic flows ${K}^{(0)}_{[0]}, {K}^{(0)}_{[1]},
{\sigma}^{(0)}_{[0]}, {\sigma}^{(0)}_{[1]}$ and the recursion
operator $\mathcal{L}$. By continuous limit one can build direct
correspondence between these flows and the continuous AKNS
isospectral and non-isospectral flows. Meanwhile, $\mathcal{L}$ goes
to the square of the AKNS recursion operartor $L_{AKNS}$ in the same
continuous limit procedure. These DAKNS flows form a Lie algebra
$\mathcal{D}$ of which the structures \eqref{K-s-Lie-akns} are
derived from the basic algebraic relations \eqref{alg-K-sigma} of
those two-potential AL flows. It has been shown that this algebra
$\mathcal{D}$ plays a key role in constructing symmetries and their
algebraic structures for both the IDNLS hierarchy and DAKNS
hierarchy.

The final main result is on the algebra deformations and explanations.
We listed out some deformations of algebras when they go to continuous case.
In fact, in Ref.\cite{Levi-TMP-01}  a contraction of subalgebras has been reported,
but in that case the correspondence between discrete case and continuous case is not direct and some linear combinations were involved.
As we can see in the present paper the correspondence between the discrete and continuous AKNS flows is direct (see \eqref{akns-order} and \eqref{akns-order-sig});
and by means of continuous limit and the lattice spacing parameter $h$
we introduced \textit{degree} for   discrete elements, as listed in \eqref{deg}.
Calculating the degree of each term one can understand the algebra deformations
before and after taking continuous limit.

Finally, we note that $\mathcal{D}$ is not a centerless KMV algebra, but it is somehow related to
this type. On one hand, $\mathcal{D}$  is derived from the centerless KMV algebra \eqref{alg-K-sigma}.
On the other hand, $\mathcal{D}$ and $\mathcal{C}$ are a continuum in continuous limit and $\mathcal{C}$
is a centerless KMV algebra.
It is not rare to see the algebraic structure changes in discrete cases. For example,
besides the subalgebra contraction found in \cite{Levi-TMP-01}
and the algebra deformations listed in this paper,
the symmetry algebra of the differential-difference KP equation also has a
non-centreless Kac-Moody-Virasoro structure\cite{zdj-DDKP-09}.
We believe  continuous limit and \textit{degree} are good means to understand these changes.

\vspace{1cm}

\section*{Acknowledgement}
This project is supported by the National Natural
Science Foundation of China (10671121) and Shanghai Leading Academic
Discipline Project (No.J50101).

\end{document}